\documentstyle[12pt]{article}
\def\be{\begin{equation}}
\def\ee{\end{equation}}

\def\ApJ{\sl Astrophys. J.}
\def\MNRAS{\sl Mon. Not. R. Astron. Soc.}

\title{\bf {Gravitational lensing constraint on the cosmic equation of state}}
\author{Deepak Jain\footnote{E--mail :
deepak@ducos.ernet.in}\,\,\,\footnote{Also at Deen Dayal Upadhyaya College,
University of Delhi, Delhi 110015},
Abha Dev, N. Panchapakesan\\
S. Mahajan and V. B. Bhatia\\
	{\em Department of Physics and Astrophysics} \\
        {\em University of Delhi, Delhi-110 007, India} 
	}

\begin {document}
\maketitle

\begin{center}
\Large{\bf Abstract}
\end{center}
\large
\baselineskip=20pt

 Recent redshift-distance measurements of Type Ia supernovae (SNe Ia) at
 cosmological distances  suggest that
 two-third of the energy density of the universe is dominated by dark
 energy
 component with an effective negative pressure. This dark energy
 component is described by the  equation of state $p_{x} = w
 \rho_{x}$ $( w \geq -1)$. We use gravitational lensing statistics
 to constrain the equation of state of this dark
 energy. We use $n(\Delta\theta)$, image separation distribution
 function of lensed quasars, as a tool to probe $w$. We find that
 for the observed range of $\Omega_m \sim 0.2 - 0.4$,   $w$ should
 lie between $-0.8 \leq w \leq -0.4$ in order to have five lensed quasars in
 a sample of 867 optical quasars. This limit is highly sensitive to
 lens  and Schechter parameters and evolution of galaxies.
   \vfill
      \eject

\baselineskip = 20pt

\begin {section} {Introduction}
Recent redshift-distance measurements of SNe Ia 
at cosmological distances  suggest that
 $2/3$ of the energy density of the universe could be in the form of a dark
 energy component with an effective negative pressure
 \cite{per,perl,re,gar}. This
 causes the universe to accelerate. There is other observational
 evidence which also supports the existence of an unknown component
 of energy density with pressure $p = w \rho$. This includes the
 recent measurements of the angular power spectrum which peak at
 $l\sim  200$ \cite{de} and the dynamical estimates of the matter density
 $\Omega_m = 0.35 \pm 0.07$ \cite{mt}. Many candidates have been
proposed  for this 
 dark energy component which is characterised by an equation of
 state $-1 \leq w \equiv p/\rho \leq 0$.

 The first candidate is the cosmological constant characterised by
 $\omega = -1$. In this case the vacuum energy density or
 the $\Lambda-\mathrm{term}\; (8\pi G\rho_{vac})$ is independent of
 time ( for a recent review see \cite{var}). There are several other
 possibilities for this dark energy:
 \begin{enumerate}

 \item A time varying  $\Lambda-\mathrm{term}$\cite{wa}

 \item Rolling scalar fields (quintessence)\cite{fri}

 \item Frustrated network of topological defects in which $w = -
 N/3$
 where N is dimension of the defect \cite{pen}.
 \item X-matter \cite{tw}

 \end{enumerate}

There are some independent methods for constraining $w$. 
For example Perlmutter et al. (1999) obtained  $w
 < -0.6$ (95\% CL) using large-scale structure and
 SNe Ia in a flat
 universe \cite{perl}. Waga \& Miceli (1999) used lensing statistics
 and  supernovae data to find that  $w <
 -0.7$ (68\% CL) in a flat universe \cite{waga}. Recently,
 Lima \& Alcaniz (2000)
 used age measurements of old high redshift galaxies (OHRG) to
 limit $w$ \cite{lima}. In flat universe for $\Omega_m = 0.3$, the
ages of OHRG give 
 $w \leq -0.27$. By combining the ``cosmic
 concordance'' method with maximum likelihood estimator, 
Wang et al. (2000) found that the
best-fit model lies in the range $\Omega_m = 0.33 \pm 0.05$ with an
 effective equation of state $w \sim -0.6 \pm 0.07$ \cite{wang}.

In this work we model this unknown form of dark energy as
X-matter which is characterized by its equation of state, $p_x =
w_x\;\rho_x$ and $w_x \geq -1$. We  use gravitational lensing
as a tool to constrain the $w$ for X-matter. In section 2 we describe
some formulae (age-redshift relationship and angular diameter
distance) used in lensing statistics. In section 3  we explain the
use of $n(\Delta\theta)$, the image separation distribution function of
lensed quasars, as a tool to constrain the cosmic equation of state
of dark energy. We  summarize our results in sec. 4.
\end {section}

\begin{section}{ Cosmology with dark energy}

 We consider spatially flat, homogeneous and isotropic cosmologies.
 The Einstein equations in the presence of nonrelativistic matter and
 dark energy are given by:
 
 \be
 H^2(z)\, \equiv \,({\dot{R}\over R})^{2}\; = \;H_{0}^{2}\left[\Omega_m
 \, ({R_0\over R})^{3} +
 \Omega_x\, ({R_0\over R})^{3(1+ w)}\right ]
 \ee
 
 \be
 ({\ddot{R}\over R}) ={-1\over 2} H_{0}^{2}\left[\Omega_m\,
 ({R_0\over
 R})^{3} +
 ( 1 + 3w)\;\;\Omega_x\, ({R_0\over R})^{3(1+ w)}\right ]
 \ee
 
 where the dot represents derivative with respect to time. Further
 $$ \Omega_m = {8 \;\pi\; G \over 3 H_0^{2}}\rho_{m0}\;\;, \;\Omega_x =
 {8 \;\pi\; G \over 3 H_0^{2}}\rho_{x0} $$ where $H_0$ is the
 Hubble
 constant at the present epoch, while $\rho_{m0} \;\;{\rm{and}}
\;\;\rho_{x0} $ are the 
 nonrelativistic matter
 density and the dark energy density respectively at the present
epoch. For more details see ref.\cite{zhu}. 
 \vskip 0.5cm
 
 The age of the universe at the redshift z is given by
 \begin{eqnarray}
 H_0\; t(z) &= &H_0\;\int_z^{\infty}{dz'\over(1 + z')H(z')}
 \nonumber \\
 & & \nonumber \\
 && =  \int_z^{\infty}{dz'\over(1 + z')\sqrt{\Omega_m(1 + z')^3 +
 \Omega_x(1 + z')^{3(1 + w)}}}. 
 \end{eqnarray}
 
The angular diameter distance between redshifts $z_1$ and $z_2$ reads as 
 \be
 d_{A}(z_1,z_2) = {R_{0}\over (1 + z_2)}\int_{z_1}^{z_2}
 {dz\over \sqrt{\Omega_m(1 + z)^3 + \Omega_x(1 + z)^{3(1 + w)}}}.
 \ee
\end{section}

\begin {section} {$n(\Delta\theta)$ as a probe}

 $n(\Delta \theta)$ is the image separation distribution function
 for lensed quasars.
 To understand $n(\Delta \theta)$ we first have 
to calculate the optical depth.
 The lensing probability or the optical depth $d\tau$ of a beam
 encountering a lensing galaxy at redshift $z_{L}$ in traversing 
 $dz_{L}$ is given by the ratio of the differential light travel
 distance $cdt$ to its mean free path between successive encounters
 with galaxies $(1/n_{L}(z)\sigma)$, $n_{L}(z)$ is the number
 density of galaxies and $\sigma$ is the effective cross-section for
 strong lensing events. Therefore,

 \begin{equation} d\tau = n_{L}(z)\sigma{cdt\over dz_{L}}
 dz_{L}.
 \end{equation}

 \noindent
 We further assume it to be conserved: $n_{L}(z) = n_{0}(1 + z_L)^{3}$. 
 The present-day galaxy luminosity function can be
 described by the Schechter function \cite{sch}
 \be
 \Phi(L, z = 0)dL \,=\,
 \phi_\ast\, \left({L \over L_\ast}\right )^\alpha \,\,\exp\left (-
 {L\over  L_\ast}\right ) \,{dL\over L_\ast}.
 \ee

 The present day comoving number density of galaxies can be calculated as

\be
 n_0 = \int_{0}^{\infty}\Phi(L)dL.
 \ee

 The Singular Isothermal Sphere (SIS) model with one dimensional velocity
 dispersion $v$ is a good  approximation to
 account for the lensing properties of a real galaxy. The deflection
  angle for all impact parameters is given by $\hat{\alpha} =
  4\pi v^{2}/c^{2}$. The lens produces two images if the angular
  position of the source is less than the critical angle
 $\beta_{cr}$,
 which is the deflection of a beam passing at any radius through an
 SIS:

 \begin{equation} \beta_{cr} =\hat{\alpha} D_{LS}/D_{OS}
 ,\end{equation}

 \noindent We use the notation $D_{OL} =
 d_A(0,z_{L}),\, D_{LS} = d_A(z_{L},z_{S}),\, D_{OS} = d_A(0,z_{S})$.
 
 \noindent
 Then the critical impact parameter is defined by $a_{cr} =
 D_{OL}\beta_{cr}$ and the cross-section is given by

 \begin{equation}
 \sigma = \pi a_{cr}^{2} = 16{\pi}^{3}\left({v \over
 c}\right)^{4}\left({D_{OL}D_{LS}\over D_{OS}}\right)^{2}.
 \label{d1}
 \end{equation}

 \noindent
 \noindent The differential probability $d\tau$ of a lensing event
 can
 be written as

 \begin{equation}
 {d\tau \over dz_{L}} = n_{L}(z)\left[ {16 \pi^{3}\over {c
 H_{0}^{3}}}
 v^{4}\,\left({D_{OL}D_{LS}\over R_0
 D_{OS}}\right)^{2}\,{1\over R_0}\right ]{cdt\over dz_{L}}dz_{L}.
 \label{d0}
 \end{equation}
 \noindent The total optical depth can be obtained by integrating
 $d\tau$ from $0$ to $z_{S}$ which is equal to {\cite{k96}}

 \begin{equation}
 \tau (z_{S}) = {F^{*}\over 30} ({D_{OS} (1 + z_{S})})^{3}
 (R_{0})^{-3}
 \end{equation}

 with  $$ F^{*} ={16\pi^{3}\over{c
 H_{0}^{3}}}\,\phi_\ast\, v_\ast^{4}\Gamma\left(\alpha +
 {4\over\gamma}
 +1\right).$$

 \noindent We neglect the contribution of spirals as lenses as their
  velocity dispersion is small as compared to ellipticals. The
 relationship between the luminosity and velocity is given by
 the Faber-Jackson relationship
 $
 {\frac{L}{L_{*}}} =
 \left({\frac{v}{v_{*}}}\right)^{\gamma}
 $.
 \vskip 0.3cm
Table 1 lists lens and Schechter parameters as given by Loveday et al. \cite{lpem}
(hereafter LPEM parameters) and Kochanek \cite{k96} 
(hereafter K96 parameters). For most of our calculations we use LPEM parameters unless
specified otherwise.

 The differential optical depth of lensing in traversing
 $dz_{L}$ with angular separation between $\phi$ and $\phi + d\phi$
 is given by \cite{ffkt}:

 \begin{eqnarray}
 {\frac{ d^{2} \tau }{ dz_{L}d\phi}}\,{d\phi}\,{dz_{L}}
 &=&{{{\gamma}/{2}} \over
 \phi} \left [{D_{OS}\over{D_{LS}}} \phi
 \right ]^{{\frac{\gamma}{2}}(\alpha + 1+ {\frac{4}{\gamma}})}
 \exp\left[-
 \,\left({D_{OS}\over{D_{LS}}} \phi\right)^{\frac{\gamma}{2}} \right
 ]{c\,dt\over dz_{L}} \nonumber \\
 & & \nonumber \\
 &&\times \, F^{*}\,{(1 + z_{L})^{3} \over \Gamma\left(\alpha +
 {4\over\gamma} +1\right)}\left[\,\left ({D_{OL}D_{LS}\over R_0
 D_{OS}}\right)^{2}\,{1\over R_0}\right ]\,{d\phi}\,{dz_{L}}
 \label{d2}
 \end{eqnarray}

 The normalized image angular separation distribution for  a source at
 $z_{S}$ is obtained by integrating the above expression over
 $z_{L}$:

 \be
 {d{\mathcal{P}}\over d\phi}\, =\, {1\over\tau(z_S)}\int_{0}^{z_{s}}\,{\frac{
 d^{2}
 \tau }{dz_{L}d\phi}} {dz_{L}}.
 \ee

 We include two correction factors in the probability of lensing:
 (1) the magnification bias and (2) the selection function due to finite resolution
 and dynamic range.

 The magnification bias $B$
 is an enhancement of the probability that a quasar is lensed. The bias
 for a quasar at redshift $z$ with apparent magnitude $m$ is written as

\be
 B(m,A_1,A_2,z) =({dN\over dm})^{-1}{ \int_{A_{1}}^{A_{2}} {dN\over
 dm}(m
 + 2.5\,logA, z)\,\, p(A)dA}
 \ee

 \noindent where $p(A)$ is the probability distribution for a
 greater
 amplification
 $A$ which is  $ 8/A^{3}$ for SIS model. We use $A_1 = 2$ and $A_2 =
 10^{4}$. We use the quasar luminosity function as
 given by Kochanek \cite{k96},
 \be
 {dN\over dm} \propto \left ( 10^{-x(m - m_0(z))} + 10^{-y(m -
 m_0(z))}\right)^{-1}
 \ee
 where
 \be
 {\rm m_0} \; (z) =
 \left\{
 \begin{array}{lll}
 m_0 + (z + 1)\; \; \; \; &{\rm if}\; \; \; & z < 1  \\
 \nonumber m_0 &{\rm if}& 1 < z < 3\\
 \nonumber m_0 - 0.7(z - 3) &{\rm if}& z > 3.
 \end{array}
 \right.
 %\label{rchi}
 \ee

 We use $ x = 1.07$, $y = 0.27$ and $m_0 = 18.92$.
 We considered a total of 862
 $(z > 1)$ highly  luminous optical quasars plus five lenses \cite{waga}.

Selection effects are caused by limitations on dynamic range,
limitations on resolution and presence of confusing sources such as
stars. Therefore we must include a selection function to correct
the probabilities. In the SIS model the selection function is modeled by
the maximum magnitude difference $\Delta m(\theta)$  that can be detected for
two images separated by $\Delta\theta$. This is equivalent
to a limit on the flux ratio $( f > 1)$ between two images $ f =
10^{0.4\,\Delta m(\theta)}$. The total magnification of images
becomes $A_{f} = A_{0}(f+1)/(f-1)$. So the survey can only detect lenses
with  magnifications larger than $A_{f}$. This sets a lower limit on the
magnification. Therefore, $A_{1}$ in the bias function gets replaced
by $A_{f}(\theta)$. To get selection function corrected
probabilities we divide our sample into two parts: the ground based
surveys and the HST Snapshot survey. We use the selection function as
suggested by Kochanek \cite{k93}.
The corrected image separation distribution function
for a single source at redshift $z_{S}$ is given as \cite{k96,chi}
\pagebreak
 \begin{eqnarray}
 P(\Delta\theta)\,& =& \, B( m , A_{f}(\theta),
 A_{2}, z){{{\gamma}/{2}} \over
 \phi} \int_{0}^{z_S}\left [{D_{OS}\over{D_{LS}}} \phi
 \right ]^{{\frac{\gamma}{2}}(\alpha + 1+ {\frac{4}{\gamma}})}
 \exp\left[-
 \,\left({D_{OS}\over{D_{LS}}} \phi\right)^{\frac{\gamma}{2}} \right
 ]\nonumber \\
 & & \nonumber \\
 &&\times \, F^{*}\,{cdt\over dz_{L}} {(1 + z_{L})^{3} \over
\Gamma\left(\alpha + 
 {4\over\gamma} +1\right)}\left[\,\left ({D_{OL}D_{LS}\over R_0
  D_{OS}}\right)^{2}\,{1\over R_0}\right ]\,\,{dz_{L}}.
 \end{eqnarray}

\noindent Similarly the corrected lensing probability for a given
source at redshift $z$ is given as \cite{k96,chi}
\be
P = \,\int {d{\mathcal{P}}\over d\phi} B( m , A_{f}(\theta),
 A_{2}, z)\; d\phi.
\ee

\noindent Here $\phi$ and ${ \Delta\theta}$ are linked through
   $\phi = {\frac{\Delta\theta}{8 \pi
 (v^{*}/c)^2}}$.

The expected number of lensed quasars is $n_{\rm L} =
\sum P_{i}$, where $P_{i}$ is the lensing probability of the ith quasar and the sum
is over the entire quasar sample. Similarly, the image-separation distribution 
function for the adopted quasar sample is $n(\Delta\theta) = \sum P_{i}( \Delta\theta)$. 
The summation is over all quasars in a given sample. 

\end {section}

\begin{section}{Results and Discussions}
Gravitational lensing statistics is a sensitive
cosmological probe for  determination of the nature of dark energy. This is
because statistics of multiply imaged lensed quasars can probe the universe
to a redshift $z \sim 1$ or even higher. This is the time when dark
energy starts playing a  dominant role in the dynamics of universe.

That lensing statistics can be used as a
tool to constrain various dark energy candidates has been known for some time.
Kochanek (1996) gave a $2\sigma$ upper bound on $\Omega_{\Lambda} < 0.66$ from
multiple  images of lensed quasars\cite{k96}. Waga \& Miceli (1999) used the
combined  analysis of gravitational lensing and Type Ia supernovae to constrain
the time  dependent cosmological term \cite{waga}.
Their combined analysis shows that $w \leq -0.7$. Cooray and Huterer (1999)
also  used lensing statistics to constrain various quintessence models
\cite{co}. However, there are  several uncertainities involved in using
gravitational lensing statistics as a tool to probe cosmology \cite{wang}.

In this article we neglect the role of spirals as
lenses as their velocity dispersion is small  as
compared to E/S0 galaxies. The constraints obtained from lensing statistics
strongly depend on lens parameters. There is a
need for updated Schechter  parameters, the faint end slope
$\alpha$ and normalisation $\phi_{*}$ for E/S0 galaxies. Earlier work on
lensing statistics used $\alpha \sim -1$, which implies the  existence of
numerous  faint E/S0 galaxies acting as lenses. Because of limited resolution,
this faint part of the luminosity function is still uncertain.
Also the parameters should be determined in a highly correlated manner from a
galaxy  survey and use of parameters derived from various surveys
might introduce error. We use the updated luminosity function
of LPEM. The LPEM luminosity function is characterised by the shallow slope
$\alpha$  at faint end and the smaller normalisation $\phi_{*}$ which shifts
the distribution to large image separations \cite{chi}. Fig. 1 shows the expected number
of lensed quasars $n(\Delta\theta  \leq 4'')$  as a function of $\Omega_{m}$ in
a flat universe with $w  = -1$ for the LPEM and K96 \cite{k96} lens and
Schechter parameters (see Table 1). We observe K96 parameters predict more
lensed quasars for this sample and for no value of the parameter $\Omega_{m}$
the expected number of lensed quasars becomes equal to 5. Moreover, as pointed out by
Chiba et al. \cite{chi} K96 parameters have been derived from various galaxy surveys and 
hence lack consistency.

Recently several galaxy surveys have come up with a much larger
sample of galaxies. This
has improved our knowledge of the galaxy luminosity function. But these surveys
don't classify the galaxies by their morphological type \cite{lc}. In lensing
statistics we need Schechter parameters of  early type galaxies  only. We feel
at the moment LPEM survey provides the most complete information regarding the
early type galaxies.

We use the  image separation distribution function
function $n(\Delta\theta)$ to constrain  the cosmic equation of state
for the dark energy.
$n(\Delta\theta)$ depends upon $w$ through the  angular diameter
distances as shown in Section 2. By varying  $w$,  the distribution
function changes which on comparison with the observations gives a
constraint on $w$.

Fig.1 shows the expected number of lensed quasars $n(\Delta\theta
\leq 4'')$  as a function of $\Omega_{m}$ in  a flat universe with $w
= -1$. Comparing the predicted numbers with the observed  lenses,
we find a value of $\Omega_{m} = 0.45$. We further generate $10^4$ data sets (quasar
sample) using the bootstrap method. We find 'best fit' $\Omega_{m}$ for each set to
obtain error bars on $\Omega_{m}$. We finally obtain $\Omega_{m} =
0.5 \pm 0.2$. Therefore the gravitational lensing statistics do not favour a
large cosmological constant.

In Fig.2  $n(\Delta\theta)$ is
plotted against $\Delta\theta $ (image separation) in the flat
cosmology for various values of $w$. The plotted rectangles
indicate the image-separation distribution of the five lensed quasars in the
optical sample  considered in this calculation.
As indicated by recent distance measurements of Type Ia supernovae,
we fix $\Omega_m = 0.3$ in Fig. 2. On comparing the theoretical prediction
of image distribution function with the observations we see  that
$w = -1$ predicts a large number of lenses which is not supported by the
observations. On the other hand, $w = 0$ gives too few  lensed
quasars. We also plot
the expected image seperation distribution  for $w =
-0.7$ and $w = -0.5$ (which corresponds to point (-0.5,0.3) on the curve in Fig. 3).
Unfortunately, we cannot do a statistical study here as the number 
of lensed quasars in this sample is only five. 	

If we increase the value of $\Omega_m$ in a  flat universe, a smaller  value
of $w$ is required to match with observations. The magnitude of the peak
$n(\Delta\theta)$  is sensitive to the value of $\Omega_m$ in a
flat universe. A larger value of $\Omega_m$ for a fixed value
of $z$ gives a smaller  angular diameter distance and hence a smaller 
value of the peak. The position of the peak of
$n(\Delta\theta)$ is sensitive to the value of $\alpha$ i.e. the faint end
slope of the luminosity function. If we take the conventional value of
$\alpha = -1$, the peak will shift to a lower value of
$\Delta\theta$ or in other words it will predict lensed images with smaller
angular separations\cite{d}.

In Fig. 3 we show the $\Omega_{m}$-$w$ plane. 
Each point on the curve corresponds
to a pair of $( \Omega_{m}, w)$ for which the expected number of lensed quasar
is equal to five ($n_{\rm L} = 5$). If the matter contribution
$\Omega_{m}$ increases, a smaller value of $w$ is required to produce
five lenses. The dotted lines in this figure correspond to the
observed range $\Omega_{m} \sim 0.2 - 0.4$\,\cite{dek}. The corresponding
range for $w$ in the flat universe is $ -0.8 \leq  w \leq -0.4$.
These results agree very well with the limit obtained on $w$  by other
independent methods  as  described in the introduction.
With this range of $w$ we can rule out atleast two dark energy
candidates. First, cosmic strings $( w \sim -0.33)$ and second the
cosmological constant $( w = -1.0)$. It is quite interesting to compare
this result with the limit obtained on $w$ by using the MAXIMA-1 AND
BOOMERANG-98 data. This data treats the dark energy as a quintessence
giving a limit $-1 \leq w
\leq -0.5$. Our constraint on $w$ is much tighter than that obtained by
using the MAXIMA-1 AND BOOMERANG-98 \cite{bal}.

The constraints obtained here depend strongly on lens and Schechter
parameters, evolution
of galaxies and, of course, on the quality of the lensing data. The
dependence of  $n(\Delta\theta)$ on lens and Schechter parameters (as 
shown in eq.(12) comes mainly through $F^{*}$ (eq.11), exponential
dependence on $\gamma$ and other related
factors present in the expression for the differential optical depth in
traversing distance $dz_L$ and with angular separation $d\phi$.
The presence of galaxy evolution decreases the optical depth and hence
the constraint on $w$ becomes weaker \cite{de1}.
The  main difficulty is that  the number of observed
lensed QSOs in this sample is too small to put
strong constraints on $w$. Extended surveys are required to
establish $n(\Delta\theta)$ as a powerful tool. The upcoming Sloan Digital
Sky Survey which is going down to $1 \sigma$ magnitude limit of $\sim
23$ will definitely increase our understanding of lensing
phenomena and  cosmological parameters. Moreover, a large number of
new gravitational lens systems $(\sim 18)$ have been discovered by the
Cosmic Lens All-Sky Survey (CLASS)\cite{tch,rt}. CLASS is the
largest survey of its kind which has more than 10000 radio sources down 
to a 5 GHz flux density of 30 mJy. But the major disadvantage of 
a radio lens survey is that there is little information on 
the redshift-dependent number-magnitude relation. This may lead to serious
systematic uncertainties in the derived cosmological constraints.

 We need a larger and complete sample of high-z QSOs, a better
 understanding of the formation and evolution of galaxies over wide
 range of redshifts and accurate luminosity function parameters before any
 definitive and strong statements can be made regarding the constraints on
 $w$ by gravitational lensing.
\end{section}

\begin{section}*{Acknowledgements}

 We would like to thank Ioav Waga for providing us with the quasar data
 and also for useful  discussions.
 \end {section}

\begin{table}
\begin{center}
\begin{tabular}{|l|lllll|}\hline\hline
$Survey$ & $ \alpha$ & $ \gamma$ & $ v^{*} (Km /s)$ & $\phi^{*}
(Mpc^{-3})$ & $ F^{*}$ \\   \hline
 \hline
 $LPEM$ & $+ 0.2$ & $4.0$ &$205$ &$3.2 \times 10^{-3}$& $0.010$   \\
 $K96$  &  $-1.0$ & $4.0$ &$225$ & $6.1 \times 10^{-3}$& $0.026$ \\
\hline
\end{tabular}
\caption{Lens and Schechter parameters for E/S0 galaxies.} 
\end{center}
\end{table}

\begin{figure}

\vskip 15 truecm

\includegraphics{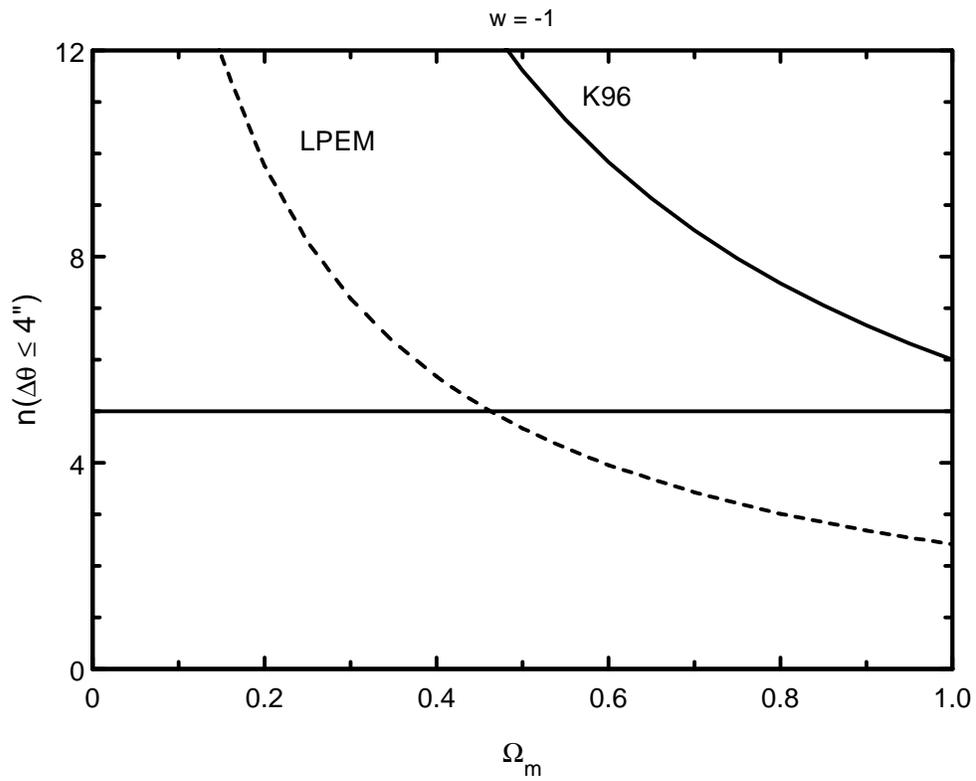}

\caption{ Predicted total number of lenses with $\Delta\theta \leq
4''$ as function of $\Omega_{m}$ for flat universe with $w = -1$
(constant $\Lambda$). The observed five lensed QSOs from optical
lens survey is shown by horizantal line.}
\end{figure}

\begin{figure}

\vskip 15 truecm

\includegraphics{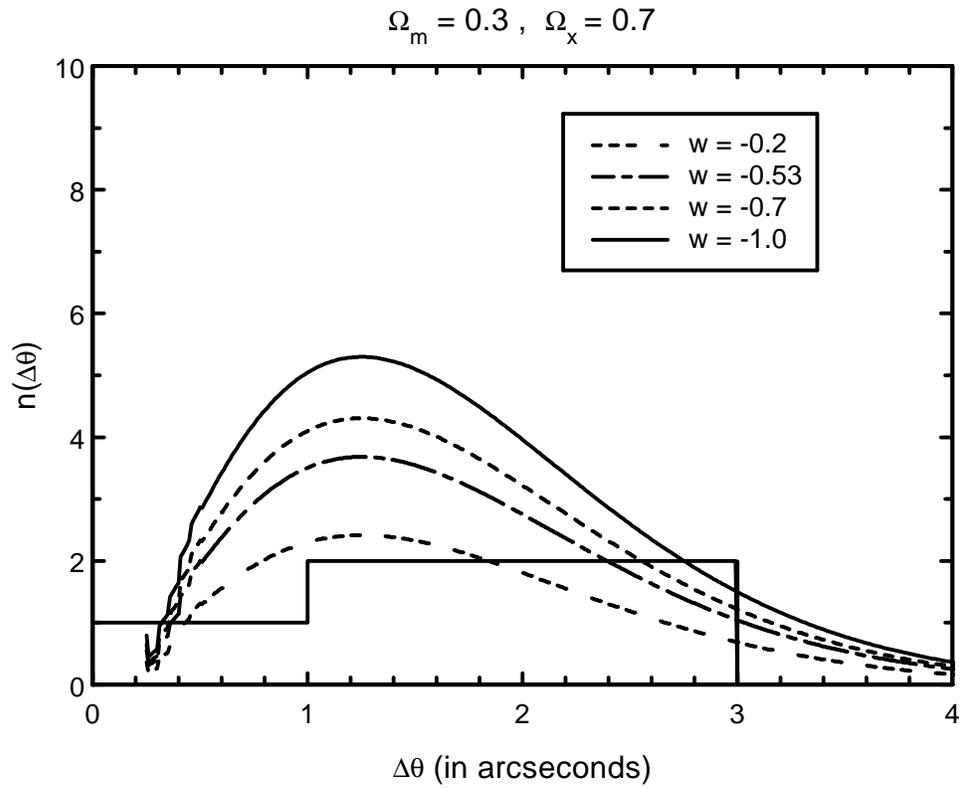}

\caption{ The expected distribution of lens image separations
with $\Omega_{m} = 0.3, \Omega_{x} = 0.7$ compared with the observed
image-separation distribution in the optical sample (histogram). The
expected distribution is plotted for different values of $w$.} 
\end{figure}

\begin{figure}

\vskip 15 truecm

\includegraphics{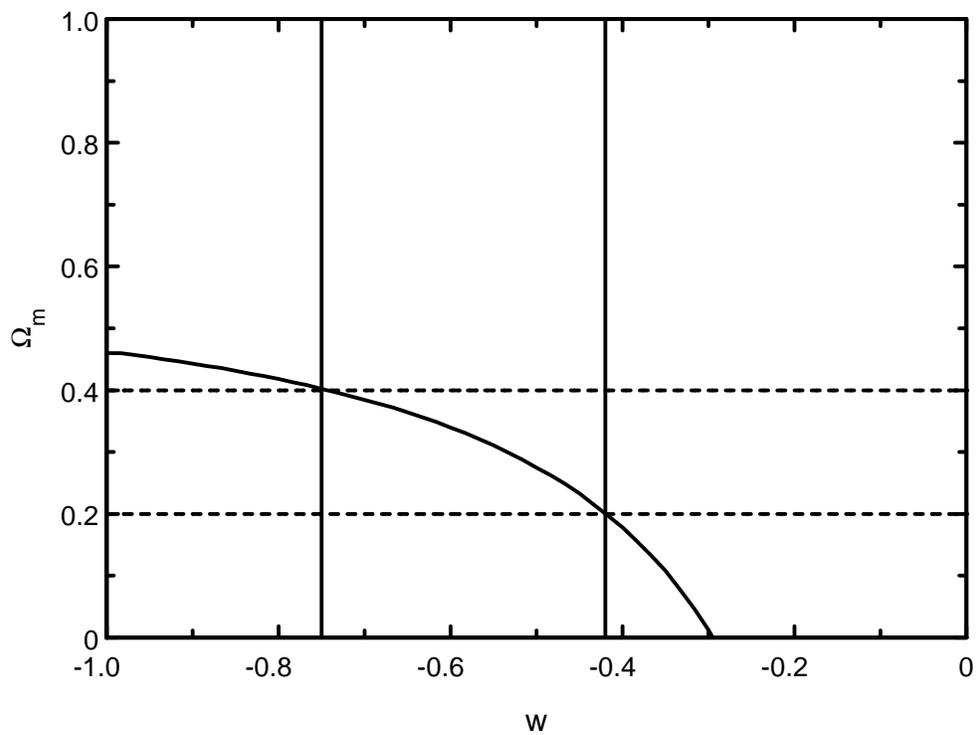}

\caption{Contour between $\Omega_m$ and $w$ in order to get
five lensed quasars in the optical sample. The dotted lines indicate
the observed range of $\Omega_m \sim 0.2 - 0.4$. The vertical lines
give the corresponding value of $w$.  }
\end{figure}

\end{document}